\documentstyle[11pt]{article}
\newcommand{\Si}{\Sigma}
\newcommand{\si}{\sigma}
\newcommand{\ove}{\overline}
\newcommand{\Sir}{\ove{\Sigma}}
\newcommand{\sir}{\ove{\sigma}}
\newcommand{\sirce}{\ove{{\sigma}}_{0}}

\newcommand{\sig}{<\!\sigma,\ove{\sigma}\!>}
\newcommand{\Sig}{\Sigma\times\ove{\Sigma}}
\newcommand{\cero}{<\!\sigma,\ove{\sigma}_{0}\!>}
\newcommand{\conu}[6]{[\hat{#1}_{#2},\hat{#3}_{#4}]=i\hbar\epsilon_{#2#4#5}
\hat{#6}_{#5}}
\newcommand{\cond}[4]{[\hat{#1}_{#2},\hat{#3}_{#4}]=0}
\newcommand{\cont}[3]{[\hat{#1}_{#2},\hat{#3}]=0}
\newcommand{\hateq}{\hat{=}}
\newcommand{\op}[2]{\hat{#1}_{#2}}
\newcommand{\lista}[2]{\newcounter{#1}\begin{list}
{$\bf #2_{\arabic{#1}}$}{\usecounter{#1}}}

\newcommand{\hs}{\hspace*{1.6em}}

\newcommand{\fa}{\forall\;}

\newcommand{\bc}{\begin{center}}
\newcommand{\ec}{\end{center}}
\newcommand{\ket}[1]{|#1\!\!>}
\newcommand{\Ket}[3]{\ket{#1 _{#2}^{#3}}}
\newcommand{\xio}[1]{{\bf A}_{\bf {#1}}}
\newcommand{\th}[1]{{\bf T}_{\bf {#1}}}

\newcommand{\defi}{\stackrel{\small Df}{=}}
\newcommand{\ai}{\'{\i}}
\newcommand{\ide}{\stackrel{id}{\leftrightarrow}}
\newcommand{\den}{\stackrel{d}{=}}

\begin{document}
\def\nobowtie{\mathrel\triangleleft\!\not\,\,\,\mathrel\triangleright}
\title{\bf Axiomatic foundations of Quantum Mechanics revisited: the case
of systems}
\author{S.E. Perez-Bergliaffa$^{1,}$ \thanks{Fellow of CONICET} ,
G.E. Romero$^{2,\:\: *}$ and H. Vucetich$^{3,}$ \thanks{Member of CONICET}}
\maketitle
\vspace*{.5cm}
\noindent$^1$ Depto. de F\'{\i}sica, UNLP, CC 67, (1900) La Plata, Argentina.\\
\noindent$^{2}$ Instituto Argentino de Radioastronom\'{\i}a, CC 5,
(1894) Villa Elisa, Bs. As., Argentina.\\
$^{3}$ FCA y G, UNLP, Paseo del Bosque S/N, (1900) La Plata,
Bs. As., Argentina.\\[0.5cm]
\begin{abstract}
 We present an axiomatization of non-relativistic Quantum Mechanics for a
system with an arbitrary number of components. The interpretation of our system
of axioms is realistic and objective. The EPR paradox and its relation with
realism is discussed in this framework. It is shown that there is no
contradiction between realism and recent experimental results.\\[0.2cm]
\end{abstract}

\section{INTRODUCTION}
\hs The interpretation of Quantum Mechanics (QM) has been a controversial
subject
over the last fifty years. A central point in this controversy is the debate
between
the realistic position and the orthodox line of the Copenhagen school. In
recent
years, this discussion has been  reheated by  some experiments
that
enabled  the testing of the implications of the paradox formulated by Einstein,
Podolsky, and Rosen (EPR) (1935). There is a widespread belief  that the
results of those experiments imply the refutation of realism and favour a
subjectivistic vision of QM. However, these conclusions originate
in an informal analysis of the structure of the theory. Any
conclusion in the aforementioned sense should be a consequence of a careful
study
of a formalized theory of QM, in such a way that all the presuppositions
and interpretation rules be explicit. Only in this case one can determine
whether the realistic interpretation of the statements is consistent with
the experimental results.

In a previous work (Perez-Bergliaffa {\em et al}. 1993), we presented a
realistic and objective
axiomatization of QM for a single microsystem from which the main
theorems can be deduced. Problems such as those arising from the EPR
paradox cannot be discussed in that axiomatic frame, because they
involve systems with more than one component. We develop here a
generalization of our preceding paper for the case of systems with an
arbitrary number of components. Armed with this new axiomatization,
we analyze some interpretational issues of QM.

We briefly present in Section 2 the ontological background of our
interpretation, because it is of the utmost importance in all our
argumentations (for details, see Bunge 1977,
1979). In Section 3 we set forth the axiomatization of the
theory, with its presuppositions, its axiomatic basis, the pertinent
definitions, and some representative theorems. In Section 4 we
discuss the relation between the EPR paradox and realism, and
then, we shortly sketch some items of the ``consistent
interpretation of QM'' that  can be deduced from our axiomatization.\footnote
{Some of the formal tools used in this work have been described
in Perez-Bergliaffa {\em et al}. (1993) (mainly mathematical tools,
such as ${\cal H}$, $G$, etc.)}

\section{ONTOLOGICAL BACKGROUND}
\hs A consistent axiomatic treatment of nonrelativistic QM
for systems with an arbitrary number of components
presuposses a {\em theory of systems}. This in turn can only be
constructed on the basis of an accurate caracterization of the
concept of {\em individual}
and its {\em properties}. In this section we caracterize a {\em physical
system}. We shall assume the realistic ontology of Bunge (a complete
and detailed analysis can be found in Bunge 1977, 1979).

The concept of individual is the basic primitive concept of any
ontological theory. Individuals associate themselves with other
individuals to yield new individuals. It follows that they satisfy a
calculus, and that they are rigorously characterized only through the
laws of such calculus. These laws are set with the aim of
reproducing the way real things associate. Specifically, it is postulated
that every individual is an element of a set $S$ in
such a way that the structure ${\cal S}=<S,\circ,\Box>$ is a
{\em commutative monoid of idempotents}. In the structure ${\cal S}$, $S$
is to be interpreted as the set of all the individuals, the element
$\Box\in S$ as the null individual, and the binary operation $\circ$
as the association of individuals. It is easy to see that there are
two classes of individuals: {\em simple} and {\em composed}.
\lista{Df}{D}
\item $x\in S$ is composed $\Leftrightarrow\exists\: y,z\in S\ni x=y\circ z$.
\item $x\in  S$ is simple $\Leftrightarrow ^{\neg}\exists\: y,z\in S\ni
x=y\circ z$.
\item $x\sqsubset y \Leftrightarrow x\circ y=y$ ( $x$ is part of $y
\Leftrightarrow x\circ y = y $).
\item ${\cal C}(x)\equiv \{y\in S\ni y\sqsubset x\}\:$(composition of $x$).
\end{list}

Real things differentiate from abstract individuals because they
have a number of properties in addition to their capability of association.
These properties can be {\em intrinsic} ($P_{\rm i}$) or {\em relational}
($P_{\rm r}$). The intrinsic properties are inherent and they are
represented by predicates or unary applications, while
relational properties are represented by n-ary predicates, with
n$>$1, as long as nonconceptual arguments are considered. For
instance, the position and the velocity of a particle are relational
properties, but its charge is an intrinsic property.

$P$ is called a {\em substantial property} if and only
if some individual $x$ possesses $P$:
\lista{Df1}{D}
\setcounter{Df1}{4}
\item $P\in{\cal P}\Leftrightarrow (\exists x)(x\in S\wedge Px)$.
\end{list}
Here ${\cal P}$ is the set of all the substantial properties.
 The set of the properties of a given individual $x$ is
\lista{Df2}{D}
\setcounter{Df2}{5}
\item $P(x)\equiv \{P\in{\cal P}\ni Px\}$.
\end{list}
If two individuals have exactly the same properties they
are the same: $\fa x,y\in S$ if $P(x)=P(y)\Rightarrow x\equiv y$.
Two individuals are identical if their intrinsic
properties are the same: $x\ide y$ ( they can differ only in their relational
properties).

A detailed account of the theory of properties is given in Bunge (1977).
We only give here two useful definitions:
\lista{Df3}{D}
\setcounter{Df3}{6}
\item $P$ is an inherited property of $x\Leftrightarrow P
\in P(x)\wedge(\exists y)(y\in{\cal C}(x)\wedge y\not= x\wedge P\in P(y))$.
\item $P$ is an emergent property of $x\Leftrightarrow P\in P(x)
\wedge(\:(\fa y)_{{\cal C}(x)}(y\neq x\:)\Rightarrow P
\notin P(y)\:)$.
\end{list}
According to these definitions, mass is an inherited property and viscosity is
an emergent property of a classical fluid.

An individual with its properties make up a {\em thing} $X$:
\lista{Df4}{D}
\setcounter{Df4}{8}
\item $X\defi <x,P(x)>$.
\end{list}
The laws of association of things follow from those of the
individuals. The association of all things is the {\em Universe} ($\si_U$).
It should not be confused with the set of all things; this is only an
abstract entity and not a thing.
Given a thing $X=<x,P(x)>$, a conceptual object
named {\em model} $X_{m}$ of the thing $X$ can be constructed by a nonempty
set $M$ and a finite
sequence {$\cal F$} of mathematical functions over $M$, each of
them formally representing a property of $x$:
\lista{Df80}{D}
\setcounter{Df80}{9}
\item $X_{m}\defi<M,{\cal F}>,$ where ${\cal F}=<{\cal F}_{1},\ldots,
{\cal F}_{n}>\ni {\cal F}_{i}:M\rightarrow V_{i}
,\: 1\leq i\leq n , V_{i} $ vector space $,{\cal F}_{i}\;
\hateq\; P_{i}\in P(x)$.
\end{list}
It is said then that $X_m$ represents $X$:  $X_{m}\hateq X$ (Bunge 1977).

The {\em state} of the thing $X$ can be characterized as follows:
\lista{Df81}{D}
\setcounter{Df81}{10}
\item Let $X$ be a thing with model $X_{m}=<M,{\cal F}>$, such that
each component of the function
\begin{center}
${\cal F}=<{\cal F}_{1},\ldots, {\cal F}_{n}>:M\rightarrow V_{1}
\times\ldots\times V_{n}$.
\end{center}
represents some $P\in P(x)$. Then ${\cal F}_{i}\:(1\leq i\leq n)$ is named
i-th state function of $X$, ${\cal F}$ is the {\em total state function }
of $X$, and the value of ${\cal F}$ for some $m\in M,\:{\cal F}(m)$,
represents the {\em state of $X$ at $m$ in the representation $X_{m}$}.
\end{list}
If all the $V_{i},\: 1\leq i\leq n$, are vector spaces,
${\cal F}$ is the state vector of $X$ in the representation
$X_{m}$, and $\:V=V_{1}\times\ldots \times V_{n}$ is the state space of
$X$ in the representation $X_{m}$.

The concept of physical law can be introduced as follows:

\lista{Df82}{D}
\setcounter{Df82}{11}
\item Let $X_{m}=<X,{\cal F}>$ be a model for $X$. Any
restriction on the possible values of the components of ${\cal F}$
and any relation between two or more of them is a {\em physical law}
if and only if it belongs to a consistent theory of the
$X$ and has been satisfactorily confirmed by the experiment.
\end{list}
We say that a thing $X$ {\em acts} on a thing $Y$ if $X$ modifies the
path of $Y$ in its space state ($X\rhd Y$: $X$ acts on $Y$).\\
We say that two things $X$ and $Y$ are {\em connected} if at least
one of them acts on the other. We come at last to the definition of
{\em system}:
\lista{Df5}{D}
\setcounter{Df5}{12}
\item A system is a thing composed by at least two connected things.
\end{list}
In particular, a physical system is a system ruled by physical laws.
A set of things is not a system, because a system is a physical entity
and not a set. A system may posses emergent properties with respect
to the component subsystems. The composition of the system $\sigma$
with respect to a class $A$ of things is (at the instant $t$):
\[C_{A}(\sigma,t)=\{X\in A\ni X\sqsubset\sigma\}\].
\lista{Df6}{D}
\setcounter{Df6}{13}
\item $\sir _{A}(\sigma ,t)=\{ X\in A\ni X\not\in {\cal C}_{A}(\sigma ,t)
\wedge(\exists Y)_{{\cal C}_{A}(\sigma ,t)}\wedge (X\rhd Y\vee Y\rhd X)\:)\}$
is the $A$-environment of $\sigma$ at $t$.
\end{list}
If $\sir _{A} (\sigma ,t)=\emptyset\Rightarrow\sigma$ is {\em closed} at the
instant
$t$. In any other case we say that it is open.\\[.5cm]
A specific physical system will be characterized by expliciting its
space of physical states. This is done in the axiomatic basis of
the physical theory. In what follows we pay particular attention to a
special type of systems: the {\em q-systems}.

\section{AXIOMATICS OF QM}
We present in this section the axiomatic structure of the theory following
the main lines of our previous paper. The advantadges of an axiomatic
formulation are discussed in Bunge (1967a).
\subsection{FORMAL BACKGROUND}
\lista{Pr}{P}
\item Ordinary bivalued logic.
\item Formal semantics (Bunge 1974a,b).
\item Mathematical analysis with its presuppositions and theory of
generalized functions (Gel'fand 1964).
\item Probability theory.
\item Group theory.
\item Association theory (Bunge 1977).
\end{list}
\subsection{MATERIAL BACKGROUND}
\lista{Prd}{P}
\setcounter{Prd}{6}
\item Cronology.
\item Physical theory of probabilities (Popper 1959).
\item Dimensional analysis.
\item Systems theory (Bunge 1977, 1979).
\end{list}

\subsection{GENERATING BASIS}
\hs The conceptual space of the theory is generated by the basis B of
primitive concepts, where

\begin{center}
B=$\{\Sir,\:$E$_{3},\:$T$,\:{\cal H}_{E},\:{\cal P},\:{\Sigma}_{1}
,\:$A, G, $\Pi,\hbar\}$
\end{center}

The elements of the basis will be semantically interpreted
by means of the
axiomatic basis of the theory, with the aid of some conventions.

\subsection{AXIOMATIC BASIS}
\hs QM is a finite-axiomatizable theory, whose axiomatic
basis is

\[
{\cal B}_{A}(QM)=\bigwedge^{36}_{i=1} {\bf A_{i}}
\]

where the index $i$ runs on the axioms.

\subsection{DEFINITIONS}
\lista{Df20}{D}
\setcounter{Df20}{14}
\item $K\defi$ set of physical reference systems.
\item $\ket{\Psi(\sigma,k)}\:\in\Psi _{\sigma}\defi$ is the representative
of the ray $\Psi_{\sigma}$ that corresponds to the system $\sigma$
with respect to $k\in K$.
\item $\Si_{N}=\Si_{1}\times\Si_{1}\times\ldots\times\Si_{1}$ ($N$ times)
is the set of all the systems composed by elements of $\Si_{1}$.\footnote{The
set $\Si_{1}$ will be characterized by $\bf A_{7}$.}
\item $\Si ^* = \{\Si _2, \Si _3,\ldots,\Si _N,\ldots\}$
\end{list}
{\bf Remark 1} With the aim of avoiding unnecesary complexity in notation we
are
not going to explicit the dependence of the operators and the eigenvalues on
the reference system.
{\bf Remark 2} The domain of the cuantified variables is explicited by
means of subindexes of the quantification parenthesis. For instance, $(\fa x)
_{A}(\exists y)_{B}(Rxy)$, means that for all $x$ in $A$, exists $y$ in $B$
such
that $Rxy$.
{\bf Remark 3} The symbol $\den$ is used for the relation of denotation (see
Bunge 1974a for details).

\subsection{AXIOMS}
\subsubsection*{GROUP I: SPACE AND TIME}
\lista{Ax}{A}
\item E$_{3}\equiv$ tridimensional euclidean space.
\item E$_{3}\:\hateq$ physical space.
\item T$\:\equiv$ interval of the real line $\cal R$.
\item T$\:\hateq$ time interval.
\item The relation $\leq$ that orders T means ``before to'' $\vee$
``simultaneous with''.
\end{list}
\subsubsection*{GROUP II: Q-SYSTEMS AND STATES}
\lista{Ax9}{A}
\setcounter{Ax9}{5}
\item $\Si_{1} ,\ove{\Si}$: nonempty numerable sets.
\item $(\forall \si)_{\Si_{1}}(\si\den$  simple microsystem).
\footnote {$\si$ satisfies the set of axioms of our previous paper, so the
definitions concerning $\si$ given there are still valid. Also note that
what is denoted here $\Si _1$ was denoted $\Si$ in that paper.}
\item $(\forall\:\si)_{{\Si}=\Si_{1}\cup\Si ^*}\:(\si\den$ q-system).\footnote{
Not every $\si \in \Si$ is necessarily a system as defined by $\bf D_{13}$.
However, with the aim of avoiding complex notation, we commit an abuse of
language in this respect.}
\item $(\forall\:\sir)_{\Sir}\:(\sir\den$ environment of some q-system).
\footnote{In particular, $\sirce\den$ the empty environment, $\cero\den$
a free q-system, and $<\!\si _{0},\ove{\si} _{0}\!>\den$ the
vacuum.}
\item ($\exists\; K)(K\subset\Sir\;\wedge$ the configuration of each $k\in K$
is independent of time).
\item $(\fa k)_{K}\:(\exists\; b)\,(\overline{b}\hateq\; k)$.
\item $(\fa\sigma)_{\Sigma}\:(\fa k)_{K}(k\nobowtie\sigma)$.
\item $(\forall \sig)_{ \Sig}\:(\exists \:{\cal H}_{E})({\cal H}_{E}
=<{\cal S},\;{\cal H},\;{\cal S'}>\:\equiv$ rigged Hilbert space).
\item There exists a one-to-one correspondence between physical states of
$\si\in\Si$  and rays $\Psi_{\sigma}\subset\:\cal H$.
\end{list}

\subsubsection*{GROUP III: OPERATORS AND PHYSICAL QUANTITIES}
\lista{Ax30}{A}
\setcounter{Ax30}{14}
\item $\cal P\equiv$ nonempty family of applications over $\Si$.
\item A $\equiv$ ring of operators over ${\cal H}_{E}$.
\item $(\forall\:P)_{\cal P} (\exists\si)_{\Si}(P\in P(\si))$.
\item ($\forall\: P)_{\cal P}(\exists\:\hat{A})_{A}\:(\hat{A}\:\hateq P)$.
\item (Hermiticity and linearity)\\[.3cm]
($\forall\:\si)_{\Si}\: (\forall\hat{A})_{A}\: (\fa P)_{\cal P}\: (\fa k)_{K}
\: (\hat{A}\hateq P\wedge\ket{\Psi (\sigma_{1},k)},\:\ket{\Psi (\sigma_{2},k)}
\:\in {\cal H}_{E}\Rightarrow$
\begin{enumerate}
\item $\ni\hat{A}[\lambda_{1}
\ket{\Psi (\sigma_{1},k)} +\lambda_{2}\ket{\Psi (\sigma_{2},k)} ]=\lambda_{1}
\hat{A}\ket{\Psi (\sigma_{1},k)}
+\lambda_{2}\hat{A}\ket{\Psi (\sigma_{2},k)}$ with $\:\lambda_{1},\lambda_{2}
\in\:{\cal C}$
\item $\hat{A}^{\dagger}=\hat{A}$).
\end{enumerate}
\item (Probability densities)\\[.3cm]
($\forall\sig)_{\Sig}\; (\forall\hat{A})_{A}\; (\fa P)_{\cal P}\;
(\fa \ket{a})_{{\cal H}_{E}}\; (\fa \ket{\Psi (\sigma ,k)})_{{\cal H}_{E}}\:
(\hat{A}\hateq P\;\wedge\;\hat{A}\ket{a}=a\ket{a}\:\Rightarrow$ the
probability density
$<\!\psi|a\!><a|\psi\!>$ corresponds to the property $P$ when
$\si$ is associated to $\sir$), that is, $\int^{a_{2}}_{a_{1}}
<\!\psi|a\!>\,<\!a|\psi\!>\,da$ is the probability for $\si$ to have
a value of $P$ in the interval $[a_{1},a_{2}]$.
\item ($\forall\:\si)_{\Si}\: (\forall\hat{A})_{ A}\:
(\fa a)_{\cal R}\: (eiv\:\hat{A}=a\wedge\hat{A}\hateq P\Rightarrow a$
is the sole value that $P$ takes on $\si$).
\item $\hbar\:\in\cal R^{+}$.
\item $[\hbar]=LMT^{-1}$.

\subsubsection*{GROUP IV: SYMMETRIES AND GROUP STRUCTURE}
\setcounter{Ax}{24}
\item (Unitary operators)\\[.3cm]
($\forall\sig)_{\Sig}\: (\forall\hat{A})_{A}\: (\fa P)_{\cal  P}\:(\fa
\hat{U}) (\hat{A}\:\hateq P\wedge\hat{U}$ is an operator on
${\cal H}_{E}\wedge \hat{U}^{\dagger}=\hat{U}^{-1}\Rightarrow\:
\hat{U}^{\dagger}\hat{A}\hat{U}\:\hateq P$).
\item $(\forall\cero)_{\Sig}\;\exists\;\hat{D}(\tilde{G}) (\hat{D}(\tilde{G})
$ is a unitary representation of rays of some central nontrivial extension
of the universal covering group $\bar{G}$ of a Lie group $G$ by an abelian
unidimensional group on ${\cal H}_{E}$).
\item The Lie algebra $\cal G$ of the group $G$ is generated by
\{$\hat{H},\;\hat{P_{i}},\;\hat{K_{i}},\;\hat{J_{i}}$\}$\:\subset A$.
\item (Algebra structure)\\[.3cm]
The structure of $\tilde{\cal G}$,
Lie algebra of $\tilde{G}$ is: \\ [.5cm]
$\conu{J}{i}{J}{j}{k}{J}\hfill\conu{J}{i}{K}{j}{k}{K}\hfill\conu{J}{i}{P}{j}{k}
{P}$ \\[.5cm]
\hspace*{\fill} $[\hat{K}_{i},\hat{H}]=i\hbar\hat{P}_{i} $\hfill
$[\hat{K}_{i},\hat{P}_{j}]=i\hbar\delta_{ij}\hat{M}\hfill $
\begin{center}
$\begin{array}{cccc}
\cont{J}{i}{H}   &\cond{K}{i}{K}{j}  &\cond{P}{i}{P}{j} &\cont{P}{j}{H}
\\[.3cm]

\cont{J}{i}{M}   &\cont{K}{i}{M}     &\cont{P}{i}{M}    &[\hat{H},\hat{M}]=0
\end{array}$
\end{center}

where $\hat{M}$ is an element of the Lie algebra of some one-parameter
subgroup (which is used to extend $\ove{G}$).
\item $\hat{M}$ has a discrete spectrum of real and positive eigenvalues.

\subsubsection*{GROUP V: GAUGE TRANSFORMATIONS AND ELECTRIC CHARGE}
\setcounter{Ax}{30}
\item$(\forall\sig)_{\Sig}\:(\fa\hat{A})_{A}(\exists\:\hat{Q})_{A}\:
(\hat{Q}\neq\hat{I}\wedge\:([\hat{Q},\hat{A}]=0)$.
\item $\hat{Q}$ has a discrete spectrum of real eigenvalues.
\item $\hat{Q}$ is the generator of gauge transformations of the
first kind.
\item There exists a sole normalizable state with
$eiv\;\hat{Q}=0$, called neutral state.
\item There exists a sole normalizable state, called vacuum, that is
invariant under $\hat{D}(\tilde{G})$ and under gauge transformations
of the first kind.

\subsubsection*{GROUP VI: COMPOSITION AXIOMS }
\setcounter{Ax}{35}
\item (Product Hilbert space)\\[.3cm]
($\forall\sig)_{\Sig}\:({\cal C}(\si)=\{\si _{1},\ldots,\si _{n}\}\Rightarrow
{\cal H}_{E}=\bigotimes_{i=1}^{n}{\cal H}_{Ei}$).
\item ($\fa\sig)_{\Sig}\:(\fa\ket{\Psi})_{{\cal H}_{E}}(\exists U_{\Pi})
(U_{\Pi}$ is a representation of a symmetric group $\Pi$ by unitary
operators $\hat{U}_{\Pi}\; \wedge$
\begin{eqnarray*}
\hat{U}_{\Pi}\ket{\Psi}&=&\hat{U}_{\Pi}\{ \Ket{\psi}{1}{a}\otimes\Ket{\psi}{2}
{b}\otimes\ldots\otimes\Ket{\psi}{n}{l}\}\\
&=&\Ket{\psi}{{\alpha}_{1}}{a}\otimes\ldots\otimes\Ket{\psi}{{\alpha}_{n}}{l}
\end{eqnarray*}
where $\{{\alpha}_{1},\ldots{\alpha}_{n}\}$ is a permutation P of $\{ 1,
\ldots n\})$.
\item $(\fa\!\sig)_{\Sig}\:(\fa\hat{A})_{A}\:
(\fa\ket{\Psi})_{{\cal H}_{E}}\: ({\cal C}(\sigma)=\{\sigma _{1},\ldots
\sigma_{n}\}\wedge\sigma _{i}\ide\sigma _{j}\wedge
\; \ket{\Psi}=\hat{U}_{\Pi}\ket{\Psi '}\:\Rightarrow\linebreak
<\Psi |\hat{A}\ket{\Psi}=<\Psi'|\hat{A}\ket{\Psi '})$.
\end{list}
{\bf Remark:} note that $\tilde{\cal G}$ is a representation by operators
of the
extended Lie algebra of the Galilei group that acts on a Hilbert space. For
other representations, see Levy-Leblond (1963).
\subsection{DEFINITIONS}
\lista{Df7}{D}
\setcounter{Df7}{18}
\item $(\fa\!\sig)_{\Sig}\:(\fa\epsilon)_{\cal R}\:(\hat{H}\ket{\Psi (\sigma
,k)}=\epsilon
\:\ket{\Psi (\sigma ,k)})
\:,\epsilon\defi$
energy of $\sigma$ in the state $\ket{\Psi (\sigma ,k)}$ with respect
to $k$ $\in$ $K$ when it is influenced by $\bar{\sigma}$.
\item $(\fa\!\sig)_{\Sig}\: (\fa p_{i})_{\cal R}\:(\hat{P}_{i}\ket{\Psi (\sigma
,k)}=p_{i}
\ket{\Psi (\sigma ,k})\:, p_{i}\defi$ i-th component of the
lineal momentum of $\sigma$ in the state $\ket{\Psi (\sigma ,k)}$
with respect to $k$ $\in$ $K$ when it is influenced by $\bar{\sigma}$.
\item $(\fa\!\sig)_{\Sig}\: (\fa j_{i})_{\cal R}\:(\hat{J}_{i}\ket{\Psi (\sigma
,k)}=j_{i}
\ket{\Psi (\sigma ,k)}),\: j_{i}\defi$ i-th component of the
angular momentum of $\sigma$ in the state $\ket{\Psi (\sigma ,k)}$
with respect to $k$ $\in$ $K$ when it is influenced by $\bar{\sigma}$.
\item $(\fa\!\sig)_{\Sig}\: (\fa m)_{\cal R}\:(\hat{M}\ket{\Psi (\sigma ,k)}=m
\ket{\Psi (\sigma ,k)})\:,m\defi$ mass of $\sigma$.
\item $\hat{X}_{i}\defi\frac{1}{m}\hat{K}_{i}$.
\item $(\fa\!\sig)_{\Sig}\: (\fa x_{i})_{\cal R}\,(\hat{X}_{i}\ket{\Psi (\sigma
,k)}=x_{i}
\ket{\Psi (\sigma ,k)})\:, x_{i}\defi$ i-th component of the position
of the {\em center of mass} of $\sigma$ in the state $\ket{\Psi (\sigma ,k)}$
with respect to $k$
$\in$ $K$ when it is influenced by $\bar{\sigma}$.
\item $(\fa q)_{\cal R}\,(\hat{Q}\ket{\Psi (\sigma,k)}=q
\ket{\Psi (\sigma ,k)})\;, q\defi$
electric charge of $\sigma$ when it is influenced by $\bar{\sigma}$.
\item ${\cal H}_{S}\equiv\{\ket{\Psi}\ni\ket{\Psi}\in {\cal H}_{E}\wedge
\hat{U}_{T}\ket{\Psi}=\ket{\Psi}$, $T$ transposition\}.
\item ${\cal H}_{A}\equiv\{\ket{\Psi}\ni\ket{\Psi}\in {\cal H}_{E}
\wedge\hat{U}_{T}\ket{\Psi}=-\ket{\Psi}$, $T$ transposition\}.
\item ${\cal H_{PS}}\defi$ space of accesible states to a given
physical system $\si \in \Si$.
\end{list}
{\bf Remark 1} The names given to the eigenvalues in the above definitions
are merely conventional and they do not imply that our axiomatics
presupposes any concept of classical physics. Any identification between
a property of the q-systems and a macroscopical property of classical physics
must be justified {\em a posteriori}. {\bf Remark 2} The meaning of the
expression {\em center of mass} can be established by means of $\bf T_1$.
\subsection{THEOREMS}
\hs In this section we give some illustrative theorems that can be
deduced from the axiomatic basis. We are not going to repeat here the theorems
valid in the case of simple microsystems ( for instance, probability
amplitudes,
Schr\"odinger equation, Heisenberg inequalities, Heisenberg equation,
superselection rules, spin). Such theorems can be found in our previous paper.
\lista{Th}{T}
\item (Additivity theorem) \\[.3cm]
$(\fa\sig)_{\Sig}\:(\fa k)_{K}\:({\cal C}(\sigma)=\{\si _{1},\ldots\si _{n}\}
\,\wedge$\\[0.2cm]
$[\hat{P}_i,\hat{X}_{jr}]=i\hbar\delta _{ij}\hfill
[\op{J}{i},\op{X}{jr}]=i\hbar\epsilon _{ijk}\op{X}{kr}\hfill
\cond{P}{i}{P}{jr}\hfill
[\op{J}{i},\op{P}{jr}]=i\hbar\epsilon _{ijk}\op{P}{kr}$\hfill \\[0.2cm]
$\cond{K}{i}{X}{jr} \hfill
[\op{K}{i},\op{P}{jr}]=i\hbar\delta _{ij}m_r$\hfill
$(i,j=1,2,3;r=1,2\ldots n)\Rightarrow$\\[0.2cm]
\hfill$\hat{P_{i}}=\sum_{s=1}^{n}\hat{P}_{is}\wedge\hat{J}_{i}=\sum_{s=1}^{n}
\hat{J}_{is}\wedge\hat{K}_{i}=\sum_{s=1}^{n}\hat{K}_{is}
\wedge\hat{M}=\sum_{s=1}^{n}\hat{M}_{s})$\hfill \\
Proof: from $\bf P_5$ and $\bf A_{35}$.
\item ${\cal H}_{S}\oplus{\cal H}_{A}\subset{\cal H}_{E}$ is a vector
subspace of ${\cal H}_{E}$.\\
Proof: from the definitions given above.
\item (Symmetrization theorem)\\[.3cm]
$(\fa\sig)_{\Sig}\:\;({\cal H}_{PS}={\cal H}_{S}\oplus{\cal H}_{A})$ .\\[.3cm]
Proof: Let $\ket{\Psi (\si)}$ be such that
\begin{equation}
\hat{U}_{T}\ket{\Psi (\si)}=\lambda _{T}\ket{\Psi (\si)}.
\end{equation}
Applying two transpositions $T_{1}$ and $T_{2}$
\[\hat{U}_{T_{1}}\hat{U}_{T_{2}}\ket{\Psi (\si)}=\lambda _{T_{1}}
\lambda _{T_{2}}\ket{\Psi (\si)}.\]
Besides, applying the transposition $T_{1}T_{2}$,
\[\hat{U}_{T_{1}T_{2}}\ket{\Psi (\si)}=\lambda _{T_{1}T_{2}}
\ket{\Psi (\si)}.\]
{}From ${\bf A_{36}}$ and ${\bf P_{5}}$, $\lambda _{T_{1}T_{2}}=\lambda
_{T_{1}}
\lambda _{T_{2}}$, and then $\lambda_{T}$ is a scalar representation
of the group $\Pi$.\\[.3cm]
There exist only two scalar representations of $\Pi$ (Cornwell 1984):
\[(\fa T)(\lambda _{T}=1)\vee (\lambda _{T}=+1,\;T\;{\rm even}\wedge
\lambda_{T}=-1,\;T\;{\rm odd}).\] Then, from (1),
\begin{equation}
\left(\ket{\Psi (\si)}\in {\cal H _{S}})\vee (\ket{\Psi (\si)}\in{\cal H_{A}}
\right).
\end{equation}
Let now be $\ket{\Psi (\si)}_{S}\,\in{\cal H_{S}}$ and $\ket{\Psi (\si)}_{A}\,
\in {\cal H_{A}}$, then,
\[_{S}<\!\!\Psi (\si)\ket{\Psi (\si)}_{A}\;=\; _{S}<\!\!\Psi
(\si)|U_{T}^{-1}U_{T}\ket{\Psi (\si)}_{A}\;=\; _{S}<\!\!\Psi
(\si)U_{T}\ket{U_{T}\Psi (\si)}_{A}= -\; _{S}<\!\!\Psi (\si)\ket{\Psi
(\si)}_{A}.\]
That is to say,
\begin{equation}
{\cal H_{A}} \perp {\cal H_{S}}
\end{equation}
Finally, from (2) and (3), ${\cal H_{PS}}={\cal H_{A}}\oplus{\cal H_{S}}$
\\[.3cm]
\noindent {\bf COROLLARY:} (Pauli's Exclusion Theorem)\\[.2cm]
$(\fa\sig)_{\Sig}\;({\cal C}(\sigma)=\{\sigma _{1},\ldots\sigma _{n}\}
\wedge\sigma _{i}\ide \sigma _{j}\Rightarrow\ket{\Psi (\sigma )}
\,\in {\cal H _{\cal PS}})$
\item ($\fa\cero)_{\Sig}\;({\cal C}(\sigma)=\{\sigma _{1}\ldots\sigma
_{n}\}\Rightarrow$
\[\hat{H}=\frac{1}{2}\sum_{i=1}^{n}\frac{\hat{p}_{i}^{2}}{m_{i}}+
\sum_{i<j} [V(r_{ij})+V(\hat{\vec{s}_{i}},\hat{\vec{s}_{j}})]\]
with
\[ V(\hat{\vec{s}_{i}},\hat{\vec{s}_{j}})=V_{1}(r_{ij})+V_{2}(r_{ij})
(\hat{\vec{s}_{i}}.
\hat{\vec{s}_{j}})+V_{3}(r_{ij})[3(\hat{\vec{s}_{i}}.\vec{n}_{ij})
(\hat{\vec{s}_{j}}
.\vec{n}_{ij})-\hat{\vec{s}_{i}}.\hat{\vec{s}_{j}}]\]
where\\[.2cm]
\hfill $r_{ij}\defi|\vec{x}_{i}-\vec{x}_{j}|$\hfill $\vec{s}_{i}
=\frac{\hbar}{2}
\vec{\tau}_{i}$\hfill $\vec{n}_{ij}\defi \frac{\vec{r}_{ij}}{r_{ij}}$
\hfill\\[.2cm] and $\vec{\tau}_{i}$ are the Pauli matrices)
\\[.3cm]
Proof: from $\xio{28}$, ${\bf P_{5}}$, and ${\bf T_{1}}$.
\end{list}
{\bf Remark 1:} The first (second)
group of commutation relations in $\bf T_1$ means that the behaviour of each
simple mycrosystem under a Euclidean motion (instantaneous Galilean
transformations) is unaffected by the presence of interactions.
{\bf Remark 2:} If $\si\in\Si$
such that $C(\si) = \{\si _1,...\si _n \}$, and $\si _i$ interacts weakly with
$\si _j$  $\Rightarrow
\hat{H}_{\si} = \sum _{i} \hat{H}_{\si _i} + O (\lambda)$, where $\lambda$ is
some coupling constant.
{\bf Remark 3:} ${\bf T_3}$ is
the so-called symmetrization postulate. Here it is a theorem implied by the
axiomatic core. {\bf Remark 4:} There exist some systems whose representative
kets have no definite symmetry when a physical space
with non-trivial topology is considered (Girardeau
1965). Such systems are excluded in the present work because of $\bf A_1$
and $\bf A_2$: it is possible to build a coordinate representation of the
operator $U_T$ in $E_3$ without any additional restriction.

\section{DISCUSSION}

\hs The axiomatization of QM in the case of a q-system with an
arbitrary number of components developed in this work is realistic and
objective. It is realistic because it assumes that the objects contained in
the ontology (that is, the set $\Si\cup\bar{\Si}$) exist independently of
sensorial experience (contrary to the fundamental thesis of
idealism). It is objective because knowing subjects or observers do not
belong to the domain of quantification of the bound variables of the
theory .

It is worth noticing that the realistic thesis does not imply that all the
functions that represent properties of real objects must have definite values
simultaneously, as classicism requires (Bunge 1989). This is clearly
seen in Heisenberg's inequalities (they follow from $\xio{28}$, see
Perez-Bergliaffa {\em et al}. 1993): they have nothing to do with measuring
devices. They reflect an inherent property of every microsystem.

At this point, an important difference should be remarked between realism and
classicism. The former is a philosophical conception regarding the nature of
the objects studied by the theory, while the latter is only a specific
feature of certain theories (see Bunge 1989).

In recent years, it has been argued that the fall of Bell's inequalities leads
to the conclusion that realism is inconsistent with experiment. However, as
we show in the next section, such a refutation does not threaten in any way
the realistic thesis adopted here.

\subsection{EPR AND REALISM}

\hs Let $\si\in\Si\ni{\cal C}(\si)=\{\si_{1}, \si_{2}\}\Rightarrow\hat{P}=
\hat{P}_{1} + \hat{P}_{2}$ by $\th{1}$. It follows from $\xio{28}$
that $[\hat{X}_{1}-\hat{X}_{2}, \hat{P}]=0$, and then,
from
$\th{8}$ of Perez-Bergliaffa {\em et al}. (1993), the quantities associated to
the operators
$\hat{X}_{1}-\hat{X}_{2}$ and $\hat{P}$ are simultaneously well-defined and
can be measured with as much precision as the state-of-the-art allows.\\
Let's suppose now that the components $\si_1$ and $\si_2$ are far away from
each other in such a way that, for the purpose of experiment, they can be
considered as isolated.  Solving Schr\"odinger's equation ($\th{4}$ of
Perez-Bergliaffa {\em et al}. 1993) in the center of mass system of
$\si$ for a null potential (see
for instance De la Pe\~na 1979), we find (in the coordinate representation)
\begin{equation}
\Psi (x_{1},x_{2})=\delta (x-a) e^{ip(x_{1}+x_{2})/2\hbar}
\label{fdo}
\end{equation}
where $a$ is the relative separation between $\si_1$ and $\si_2$. If we now
measure the position of $\si_1$ we can infer (from the relation $x_{1}-x_{2}
=a$) which value would be found if we measure the position of $\si_2$
immediately after the first measure has been carried out. Assuming that there
is no action-at-distance in a quantum sense (i.e. that two subsystems
apart enough from each other can be considered as isolated, an assumption
known as locality or separability), the
inference of $x_2$ is made without perturbing $\si_2$ in any way. It follows
then that the position of $\si_2$ has a definite predetermined value not
included in (4). This implies that the description given by QM
is incomplete. By the same reasoning, it can be inferred that the
lineal
momentum of $\si_2$ has also a definite value, at variance with Heisenberg's
inequalities.
Then both the position and the lineal momentum of $\si_2$ have a definite
predetermined value: we do not have to work out any additional measure to
know them. This clearly contradicts the subjetivistic interpretation of
Copenhagen.

The argument given above is a brief account of the so-called ``EPR paradox''.
In short, it
states that if locality is accepted in QM then the
theory must be incomplete. In other words, the theory must have hidden
variables (Bohm 1953). Besides, a theorem due to Bell (1966)
shows that the predictions of deterministic, local theories that have hidden
variables can be compared, by means of a given class of experiments, with the
predictions of QM. Experiments of such a class have been
carried out by Aspect {\em et al}. (1991, 1992), and their results are in
complete agreement with QM.

The reader should note that these results do not affect the realistic
philosophy that underlies our axiomatization. In fact, as it was shown by
Clauser and Shimony (1978),

\begin{center}
(Hidden Variables $\wedge$ Separability) $\Rightarrow$ (Bell's inequalities)
\end{center}

It follows that if Bell's inequalities are refuted by recourse to the
experiment, then (1) theories with hidden variables are false (i.e. QM is
complete) or (2) the theory is non-local or (3) both (1) and (2) are true. The
axiomatization we present here assumes non-locality and completeness, so it
{\em predicts} that Bell's inequalities are false. The non-locality
originates in the systemic point of view adopted in the background material
(more precisely, in ${\bf P}_{\bf 10}$; see Section 2 for details), while
completeness is introduced through $\xio{19}$, according which every
property of the physical system under study has its mathematical counterpart
uniquely defined in the theory.

In brief, the axiomatization we present here is realistic, objective,
non-local, and complete. These features are essential for the study
of quantum cosmology, a subject in which the orthodox (subjetivistic)
interpretation cannot be applied succesfully.

The system formed by the association of all the things is the Universe
($\si_U$, see Section 2). By definition, the environment of $\si_U$ is the
empty environment: $\bar{\si}_U = \sirce$. It follows that any
interpretation of QM that requires external observers to produce the
collapse of the wave function cannot be applied to the study of $\si_U$.
In this case
it is mandatory to have at our disposal an objective interpretation. The usual
approach (based on the wave function) presuposses the interpretation of
Everett (1957) or variations of it (see for instance Halliwell 1992).
Our axiomatization shares
with Everett's interpretation the realism and the needless of
Von Neumann's projection postulate. However, the theory of measurement that
follows from our axiomatization does not entail the introduction of the
``Many Worlds'', as will be discussed elsewhere.

\subsection{SOME REMARKS ON THE ``CONSISTENT INTERPRETATION''}

\hs Recently, Griffiths (1984), Omn\`es (1992), and Gell-Mann and Hartle (1990)
have developed a new
formulation of QM: the so-called  ``consistent
interpretation''. They claim it is both realistic
and objective. In the following, we shall argue that
their main physical results can be obtained as theorems in our
formalism, although detailed proofs, which are lengthy, will be
presented elsewhere.

In the consistent interpretation, the density matrix plays a central role.
This concept is secondary in our
axiomatization because the notion of partition of a system $\sigma$
in two subsystems (i.e. $\sigma=\sigma_1\dot{+}\sigma_2$, where the symbol
$\dot{+}$ means physical sum, see Bunge 1967b) has been incorporated to the
ontological background.
Starting from this partition, it is possible to show that the state of each
subsystem is represented by a density operator $\rho_i$ (see
Balian 1982 for a nonrigorous proof).

The existence of the classical limit can be proved in our
formulation essentially in the same way as in Omn\'es (1992).
Specifically, there exists a many-to-one partial
function ${\cal C}$ that associates a function ${\cal A}(p,q)$ (that
depends on classical phase space variables)
to operators $\hat{A}(\hat{p},\hat{q})$. The function ${\cal A}(p,q)$
is the classical counterpart of $\hat{A}(\hat{p},\hat{q})$.
The function ${\cal C}$ is many to one because, due to
the lack of commutativity of the operator ring, several operators have
the same classical counterpart, and it is partial
because dynamical variables such as spin have no classical counterpart.

With these elements and the aid of our axiomatics, we could construct a
``theory of measurement''. If the system $\sigma$ is decomposed
as follows:
\begin{equation}
\sigma = \sigma_S \dot{+} \sigma_A \dot{+}\bar{\sigma}_S
\end{equation}
where $\sigma_S$ is the subsytem on which the measure is
performed, $\sigma_A$ is the
``apparatus'' and $ \bar{\sigma}_S$ is the ``environment'', then, with
suitable restrictions on the three subsystems, the main results
of measurement theory could be deduced as in Omn\`es (1992).\footnote{We should
remark that the resulting measurement theory does not apply to real situations
but to the analysis of highly idealized typical experiments: it can predict
accurately no outcome of a single real experiment (Bunge 1967b)}

``Wave packet reduction'' can be expressed as a trace on the
density matrix of the ``apparatus'' subsystem
(L\"uders 1951, Omn\`es 1992). This is probably the closest one can get
to a
proof of ``von Neumann's projection postulate'' in our formulation.
However, no physical
process is involved in the reduction: it is a mathematical device to
describe a subset of initial conditions (Omn\`es 1992).
\footnote{The main role of the environment is to
produce decoherence on the density matrix of the other two subsystems,
forcing them into a diagonal form (Omn\'es 1992, Paz 1994). There should
exist a many-one function mapping (sets of) states of $\Sigma_S$ into
well defined states of $\Sigma_A$.}

\section{CONCLUDING REMARKS}

\hs We finally would like to point out here that certain realistic
interpretations of QM cannot
face succesfully the refutation of Bell's inequalities. This is true for
deterministic interpretations, i.e.  interpretations that imply the existence
of hidden variables that complete the classical characterization of the state
of the particles that compose the statistical {\em ensembles}. This failure is
avoided by a literal (i.e. strictly quantum) interpretation. We have shown
here that such an interpretation is possible. Moreover, our axiomatics
offers a well-suited frame for the analysis of recent attempts focused on
obtaining the classical limit as an emergent property in a macroscopical
system from the constituent microsystems, by means of a decoherence
process. This line of research will be developed elsewhere.

\section*{ACKNOWLEDGEMENTS}
\hs It is a pleasure to thank Prof. M. Bunge for encouragement and valuable
suggestions. This work has been partially supported by CONICET and the
National University of La Plata.

\section*{REFERENCES}

{\sc Aspect, A., Grangier, P., and
Roger, G.} (1981). {\em Phys. Rev. Let.} {\bf 47}, 460.\\
{\sc Aspect, A., Dalibard, J., and
Roger, G.} (1982). {\em Phys. Rev. Let.} {\bf 49}, 1804.\\
{\sc Balian, R} (1982). {\em Du Microscopique au Macroscopique: Cours de
Physique
Statistique de l'Ecole Polytechnique}, Ec\`ole Politechnique, Paris.\\
{\sc Bohm, D.} (1953). {\em Quantum Theory}, Prentice-Hall, Englewood Cliffs,
N. J.\\
{\sc Bell, J.} (1966). {\em Rev. Mod. Phys.} {\bf 38}, 447.\\
{\sc Bunge, M.} (1967a). {\em Rev. Mod. Phys.} {\bf 39}, 463.\\
{\sc Bunge, M.} (1967b). {\em Foundations of Physics}, Springer-Verlag,
New York.\\
{\sc Bunge, M.} (1974). {\em Sense and Reference}, D. Reidel Publ. Co.,
Dordrecht and Boston.\\
{\sc Bunge, M.} (1974). {\em Interpretation and Truth}, D. Reidel Publ. Co.,
Dordrecht and Boston.\\
{\sc Bunge, M.} (1977). {\em The Furniture of the
World}, D. Reidel Publ. Co., Dordrecht and Boston.\\
{\sc Bunge, M.} (1979). {\em A World of Systems},
D. Reidel Publ. Co., Dordrecht and Boston.\\
{\sc Bunge, M.} (1989). {\em Phil. Nat.} Band 26, Heft 1, 121. \\
{\sc Clauser, J. and Shimony, A.}
(1978). {\em Rep. Prog. Phys.} {\bf 41}, 1881.\\
{\sc Cornwell, J}. (1984). {\em Group Theory in Physics}, Academic Press,
New York.\\
{\sc De la Pe\~na, L.} (1979) {\em Introducci\'on a la Mec\'anica Cu\'antica}
, Compa\~n\ai a Editorial Continental, M\'exico.\\
{\sc Einstein, A., Podolsky, B., and
Rosen, N.}, (1953). {\em Phys. Rev.} {\bf 47}, 777.\\
{\sc Everett, H.} (1957). {\em Rev. Mod. Phys.} {\bf 29}, 454.\\
{\sc Gel'fand, I. M and Shilov, G. E.} (1967).
{\em Generalized Functions}, vol. 3, Academic Press, New York.\\
{\sc Gel'fand, I. M et al} (1964-1968).
{\em Generalized Functions},vol. 1-4, Academic Press, New York.\\
{\sc Gell-Mann, M. and Hartle, J.~B.} (1990).
{\em Proc. of the 3rd International Symposium on Foundations of Quantum
Mechanics in the Light of New Technology}, ed. by S. Kobayashi, H. Ezawa,
Y. Murayama and S. Nomura, Physical Society of Japan, Tokyo.\\
{\sc Griffiths, R.} (1984). {\em J. Stat. Phys.} {\bf 36,} 219.\\
{\sc Halliwell, J.} (1992). {\em Proceedings of the
13th International Conference in General Relativity and Gravitation}, IOP
Publishing, C\'ordoba, Argentina .\\
{\sc Levy-Leblond, J-M.} (1963). {\em J. Math, Phys.} {\bf 4}, 776.\\
{\sc L\"uders, V. G.} (1951). {\em Ann. Phys. (Leipzig)} {\bf 8,} 322.\\
{\sc Omn\`es, R.} (1992). {\em Rev. Mod. Phys.} {\bf 64,} 339.\\
{\sc Paz, J.~P.} (1994). To be published in the {\em Proceedings of the
$6^\circ$ National Symposium on Relativity and Gravitation,} C\'ordoba,
Argentina. \\
{\sc Perez-Bergliaffa, S.E.,
Romero, G.E. and Vucetich H.} (1993). {\em Int. J. Theoret. Phys.} {\bf 32},
1507.\\
{\sc Popper, K. R.} (1959). {\em Brit. J. Phil. Sci.} ${\bf 10}$, 25.\\

\end{document}